# Complex Effects of Salt on Small-Angle X-ray Scattering of BSA Originate From the Interplay of Ions and Hydration Water

Anshika Dhiman[1], Sanbo Qin[1], and Huan-Xiang Zhou[1,2]*

[1]Department of Chemistry and Department of Physics, University of Illinois Chicago, Chicago, Illinois 60607, USA

*Correspondence: hzhou43@uic.edu

ABSTRACT

Salts are an integral part of the environment for living systems and, therefore, understanding their effects on proteins and other biomolecules is of fundamental interest. Small-angle X-ray scattering (SAXS) of protein solutions can provide valuable information on salt effects, but extracting this information has been a significant challenge. For example, SAXS data of bovine serum albumin (BSA) at various salt concentrations were fit to three different spherical models. Here we combined the newly developed FMAPIq approach with explicit-solvent all-atom molecular dynamics simulations to show that the complex effects of salt on the SAXS of BSA originate from the interplay of ions and hydration water, leading to a general picture of protein-ion-water interactions.

In addition to water, salts are ubiquitous in the environments of living systems and within individual cells. Salts can exert complex effects on protein folding, binding, and various forms of condensation including crystallization, precipitation, and liquid-liquid phase separation [1-14]. Salt ions can both directly bind to protein residues and indirectly modulate intra- and inter-protein interactions by various means, e.g., screening electrostatic interactions, strengthening hydrophobic interactions, and perturbing hydration water. Given the complexity of salt effects, it is rare to achieve a unified understanding with atomic-level insight. Here we report our work in achieving such an understanding for the salt effects on the small-angle X-ray scattering (SAXS) of bovine serum albumin (BSA).

SAXS can provide valuable information on inter-protein interactions and salt effects. However, extracting this information has posed a significant challenge. The common practice in interpreting SAXS data is to fit with spherical models, in which inter-protein interactions are simplified to only depend on the center-to-center distance. Salt effects are then accounted for implicitly by varying parameters in the spherical models. Zhang et al. [7] collected exquisite SAXS data of 100 mg/mL BSA over NaCl concentrations from 0 to 2 M. When fitting their data to spherical models, they had to use three different models. At low salt (0 to 0.05 M), the data were fit to a hard-sphere model plus screened electrostatic repulsion. At intermediate salt (0.1 to 0.5 M), the fit to the screened electrostatic model was no longer satisfactory, as the required net charge was excessive. A fit to a pure hard-sphere model was also attempted, although the required protein volume fraction was higher than expected. Lastly, the data at high salt (1 to 2 M) were fit to a hard-sphere model plus a square well.

To overcome the limitations of spherical models, we recently developed an approach called FMAPIq that combines atomistic simulations of protein solutions (MAEPPIsim) and direct calculations of SAXS [i.e., $I(q)$] profiles on snapshots from the simulations (Sim2Iq) [15]. MAEPPIsim was made possible by precomputing protein-protein pair interactions. Sim2Iq was implemented by calculating the scattering amplitude as the Fourier transform of the excess electron density. The latter consists of three terms:

$$\Delta\rho(\mathbf{r}) = \rho_{\text{vac}}(\mathbf{r}) - \rho_{\text{sol}}(\mathbf{r}) + \Delta\rho_{\text{hyd}}H(\mathbf{r}) \tag{1}$$

which represent the electron density of the protein molecules when placed in vacuum, the electron density of the would-be bulk solvent in the region excluded by the protein molecules, and the difference in electron density between hydration and bulk solvent in the hydration shell [represented by the function $H(\mathbf{r})$], respectively. The default values of $\rho_{\text{sol}}$ and $\Delta\rho_{\text{hyd}}$ were 0.334 and 0.011 e/Å$^3$, respectively. FMAPIq calculations revealed artifacts generated by spherical models and accurately reproduced experimental results, leading to sound physical interpretations, including dimer formation by BSA at high concentrations. However, the BSA calculations were limited to low salt.

In Figure 1a, we compare the $I(q)$ profiles of 100 mg/mL BSA over NaCl concentrations from 0 to 2 M calculated by FMAPIq and their experimental counterparts by Zhang et al. [7]. $I(q)$ is presented after normalization by the protein number density ($n_P$). As shown previously, the FMAPIq results match well with the experimental data at 0 NaCl [15]. The match is still reasonable at 0.05 M but deteriorates quickly at higher salt. In particular, the FMAPIq results saturate above 0.5 M NaCl, as electrostatic repulsion between the highly negatively charged BSA molecules is screened out at high salt, but the experimental $I(q)$ data at low $q$ show a highly nonmonotonic behavior, first increasing from 0 to 0.05 M NaCl, then decreasing from 0.05 M to 0.1 M NaCl but increasing again from 0.1 to 1 M NaCl, and lastly decreasing from 1 M to 2 M NaCl.

We first recognized that the electron density, $\rho_{\text{sol}}$, of the bulk solvent depends on NaCl concentration. For example, using its density (0.9982 g/cm$^3$) at 20 ºC, the electron density of pure water can be calculated to be 0.3337 e/Å$^3$. At 1 M NaCl, the density of the solvent becomes 1.0382 g/cm$^3$ [16], and the corresponding electron density is 0.3443 e/Å$^3$. For up to 2 M NaCl, the solvent electron density has a linear dependence on NaCl molarity ([NaCl]): $\rho_{sol} = 0.3338 + 0.0105[\text{NaCl}]$. Accounting for the salt dependence of $\rho_{sol}$, the FMAPIq results move closer to the experimental data (Figure 1b, c). The match now becomes very good at 1 M NaCl, but the experimental $I(q)$ values at low $q$ are still overestimated at 0.1 and 0.5 M and then again at 1.5 and 2 M NaCl.

We suspected that the hydration-shell contrast, $\Delta\rho_{hyd}$, was overestimated by the default value, 0.011 e/Å$^3$. This value is supposed to be used in combination with the default bulk solvent electron density $\rho_{sol}$. Because $\rho_{sol}$ is now elevated by NaCl, the shell contrast may not be as high as the value relative to pure water. By reducing $\Delta\rho_{hyd}$ from 0.011 e/Å$^3$, FMAPIq results can match well with the experimental data at all [NaCl] (Figure 1d, e). The reductions are not required at 0 and 1 M NaCl, slight (to 0.010 e/Å$^3$) at both 0.05 and 1.5 M NaCl, and more substantial at 0.1 and 0.5 M NaCl (to 0.002 and 0.007 e/Å$^3$) and at 2 M NaCl (to 0.008 e/Å$^3$). We display the total electron density ($\rho_{hyd} \equiv \rho_{sol} + \Delta\rho_{hyd}$) in the hydration shell in Figure 1f (solid circles), which exhibits a significant dip at 0.1 M NaCl.

To explain the nonmonotonic dependence of $\rho_{hyd}$ on [NaCl], we performed all-atom explicit-solvent molecular dynamics (MD) simulations of a BSA molecule in the presence of the full range of [NaCl]. Figure 2a, b presents the number of Na$^+$ and Cl$^-$ ions and water molecules in the hydration shell of BSA as a function of [NaCl]. As expected, the ion counts increase monotonically with [NaCl]. Interestingly, the water count is nonmonotonic, with a dependence on [NaCl] similar in shape to that exhibited by $\rho_{hyd}$ derived from FMAPIq (Figure 1f, solid circles), including a minimum at 0.1 M NaCl. Using the ion and water counts, we calculated the total electron density in the hydration shell. The results, shown as open circles in Figure 1f, match closely with those derived from FMAPIq. That is, the explicit-solvent MD simulations produce hydration-shell electron densities needed by FMAPIq to fit the experimental SAXS data of Zhang et al., and the nonmonotonic behavior of $I(q)$ can be traced to the water count in the hydration shell of BSA.

Why does the water count in the hydration shell have a nonmonotonic dependence on [NaCl]? To begin addressing this question, we notice that the increase in ion counts is concave at low [NaCl] and then becomes linear at higher [NaCl] (Figure 2a). These two features suggest two ion populations (Figure 2c): those tightly bound to sites on the protein surface, with a total number $N_{\text{ion}}^{\text{t}}$ that is saturable, and those loosely bound, with a total number $N_{\text{ion}}^{\text{l}}$ that is proportional to the bulk concentration [NaCl]. The number of Na$^+$ or Cl$^-$ in the hydration shell is thus

$$N_{\text{ion}} = \frac{N_{\text{ion}}^{\text{tt}}}{1 + K_{\text{ion}}/[\text{NaCl}]} + \alpha[\text{NaCl}]V_{\text{hyd}} \qquad (2)$$

where "ion" = Na$^+$ or Cl$^-$, $N_{\text{ion}}^{\text{tt}}$ is the total number of tightly bound ions at saturation, $K_{\text{ion}}$ is the dissociation constant for tight binding, $V_{\text{hyd}}$ is the volume of the hydration shell, and $\alpha$ is a constant of order 1. Fitting the ion counts to Eq (2) (Figure 2a, dotted curves) yielded $N_{\text{Na}^+}^{\text{tt}}$ = 22.3, $K_{\text{Na}^+}$ = 0.21 M, $N_{\text{Cl}^-}^{\text{tt}}$ = 4.9, $K_{\text{Cl}^-}$ = 0.04 M.

The initial decrease in water count occurs at low [NaCl] where the tightly bound ion population dominates; these ions displace water molecules that would otherwise coordinate the protein sites [Figure 2d, step labeled as (1)]. At intermediate [NaCl], the loosely bound ion population takes over, which draws water into the hydration shell [Figure 2d, step (2)], thereby countering the action of the tightly bound ion population. At high [NaCl], the coordination shells of loosely bound ions overlap, and therefore the number of water molecules, on a per ion basis, drawn into the hydration shell of BSA is reduced [Figure 2d, step (3)]. These considerations lead to the following expression for the water count in the hydration shell of BSA:

$$N_{\text{water}} = N_0 - A_{\text{Na}^+}N_{\text{Na}^+}^{\text{t}} - A_{\text{Cl}^-}N_{\text{Cl}^-}^{\text{t}} + BN^{\text{l}} \qquad (3)$$

where $N_0$ is the water count in the total absence of ions, $A_{\text{ion}}$ is the number of water molecules displaced by each tightly bound ion, $N^{\text{l}}$ is the total number of loosely bound

Na$^+$ and Cl$^-$ ions, and $B$ is the number of water molecules drawn into the hydration shell of BSA by each loosely bound ion. The latter is a decreasing function of $N^l$:

$$B = \begin{cases} B_0, & N^l < \dfrac{B_1 - B_0}{B_2} \\ B_1 - B_2 N^l, & N^l \geq \dfrac{B_1 - B_0}{B_2} \end{cases} \quad (4)$$

The fit of the MD data to Eqs (3) and (4) are displayed as a solid curve in Figure 2b, yielding 4 or 0.8 water molecules displaced per tightly bound Cl$^-$ or Na$^+$ ion, 2 water molecules drawn in per loosely bound ion at NaCl up to 0.5 M NaCl. The latter value tapers off at higher NaCl, to 0.03 water molecules per ion at 2 M NaCl.

To validate the foregoing interpretation of the interplay between ions and hydration water, we examined possible correlations between hydration water count and ion counts in the individual simulations. We expect to observe a negative correlation between water count and Cl$^-$ count at low NaCl where the tightly bound population dominates, but a positive correlation at high NaCl where the loosely bound population dominates. Such opposite correlations are indeed observed (Figure 2e).

Lastly we present the atomistic picture of ions and water in the hydration shell (Figure 3a). We defined tightly bound ions as those coordinated with at least two protein N or O atoms; the resulting counts of tightly bound ions agree with those from fitting to Eq (2) (Figure 3b). Figure 3c (right panel) illustrates two tightly bound Cl$^-$ ions at 0.1 M NaCl, one coordinated with R194 and R198 and often also with W213, the other coordinated with R217 and R256 and often also with H241. At 0 M NaCl, these sidechains are solvated by water (Figure 3c left panel); in the presence of NaCl, some of these water molecules are thus displaced by the two tightly bound Cl$^-$ ions. Tightly bound Na$^+$ ions are usually coordinated with multiple Asp and/or Glu residues (Figure 3d). Loosely bound ions are predominantly coordinated by water (Figure 3e).

We quantified the BSA partner groups of ions and water by calculating the latter's radial distribution functions (Figure 4, top panels). The results reveal that, on a per residue basis, Asp and Glu carboxyls are the most frequent partners of $Na^+$ ions, followed by Asn and Gln side-chain carbonyls and then by Ser and Thr hydroxyls and backbone carbonyl. For $Cl^-$ ions, the most frequent partners are Arg guanidium and Lys amine, followed by Asn and Gln side-chain amine and Ser hydroxyl and then by backbone amide. All these partner groups of ions are also frequent partners of hydration water. When accumulated over all residues of a given type, Glu, Asp, and backbone carbonyl interact with 21, 18, and 14 $Na^+$ ions, respectively, while Lys and Arg interact with 10 and 5 $Cl^-$ ions, respectively (Figure 4, bottom panels) at 1 M NaCl. Interestingly, the top partner groups of ions identified here for a structured protein are the same as those found for a disordered protein in a dense phase [13], indicating that the picture of protein-ion-water interactions delineated here is general.

This work was supported in part by Grant GM118091 from the National Institutes of Health.

FIGURE CAPTIONS

FIG. 1. SAXS profiles of BSA at pH 7 and 20 °C over a wide range of NaCl concentrations. (a) Comparison of experimental data (dashed curves) from Zhang et al. [7] and those calculated by FMAPIq (solid curves) using the default solvent electron density and hydration-shell contrast. The legend lists [NaCl] in M. FMAPIq calculations were as described previously [15], with 10% of BSA molecules modeled as dimers and ionic strength as $I_c$ = [NaCl] + 0.02 M. (b, c) Same as (a) but with the solvent electron density given by $\rho_{sol} = 0.3338 + 0.0105 I_c$. (d, e) Same as (b, c) but with adjusted hydration-shell contrast. (f) Total electron density, $\rho_{hyd}$, in the hydration shell. Filled circles: sum of $\rho_{sol}$ and $\Delta\rho_{hyd}$ in (d, e); open symbols: calculated from ion and water counts in Figure 2a, b.

FIG. 2. Ion and water counts in the hydration shell of BSA from all-atom explicit-solvent MD simulations. (a) $Na^+$ and $Cl^-$ counts. Mean values and standard deviations among four replicate MD simulations are displayed as circles and error bars; fits to Eq (2) are presented as dashed curves. For $Na^+$, 4.8 ions (from 17 neutralizing ions) were bound to BSA at 0 NaCl; this number was added to the right-hand side of Eq (2) and counted toward the total number of tightly bound $Na^+$. (b) Water count. MD data are displayed as circles connected by dashed lines [colors of circles at four NaCl concentrations are changed to match those in (e)]; fit to Eqs (3) and (4) is presented as a solid curve. For (a, b), MD simulations were performed using GROMACS 2023.1, with the AMBER99SBws force field for the protein [17], the TIP4P/2005s model for water [18], and associated parameters for ions. The system was set up with the crystal structure (Protein Data Bank entry 3V03) solvated in a cubic box with a 140-Å side length, neutralized with 17 $Na^+$ ions, and NaCl added to the desired concentrations. Four replicate simulations were run for 1000 ns and analysis was done on the last 500 ns. The hydration shell was defined as 4 Å (see Figure 4 top panels) within BSA heavy atoms. (c) Two populations of ions in the hydration shell. (d) Interplay of ions and water in the hydration shell. (1) At low NaCl, tightly bound ions

displace hydration water; at intermediate NaCl, loosely bound ions draw water into the hydration shell; at high NaCl, coordination shells of loosely bound ions overlap with each other and hence the number of water molecules, on a per ion basis, drawn into the protein hydration shell is reduced. (e) Negative and positive correlations between hydration-shell $Cl^-$ and water counts in the four replicate simulations at the indicated [NaCl].

FIG. 3. Characteristics of tightly and loosely bound ions. (a) A snapshot of BSA simulations at 0.1 M NaCl. Hydration water is shown as thin sticks; $Cl^-$ and $Na^+$ ions are shown as red and blue spheres, respectively; side chains that coordinate ions are shown as thick sticks. (b) Comparison of expected (curves) and actual (symbols) counts of tightly bound ions. Error bars represent standard deviations among four replicate simulations. (c) Two tightly bound $Cl^-$ ions displace water. Left panel: from a simulation at 0 NaCl; right panel: zoomed view of a boxed region in (a). (d) Zoomed view of a boxed region in (a), illustrating a tightly bound $Na^+$ ion. (e) Zoomed views of loosely bound $Cl^-$ and $Na^+$ ions.

FIG. 4. Radial distribution functions and number of bound ions and water at 1 M NaCl. (a) $Na^+$. (b) $Cl^-$. (c) Water. Top panels: radial distribution functions; a dashed vertical line is drawn at 4 Å to indicate the cutoff for defining the hydration shell. Bottom panels: average number of ions or water bound to a single residue of a given type (bars in darker color, left ordinate) or average total number of ions or water bound to all residues of a given type (bars in lighter color, right ordinate). Error bars represent standard deviations among four replicate simulations.

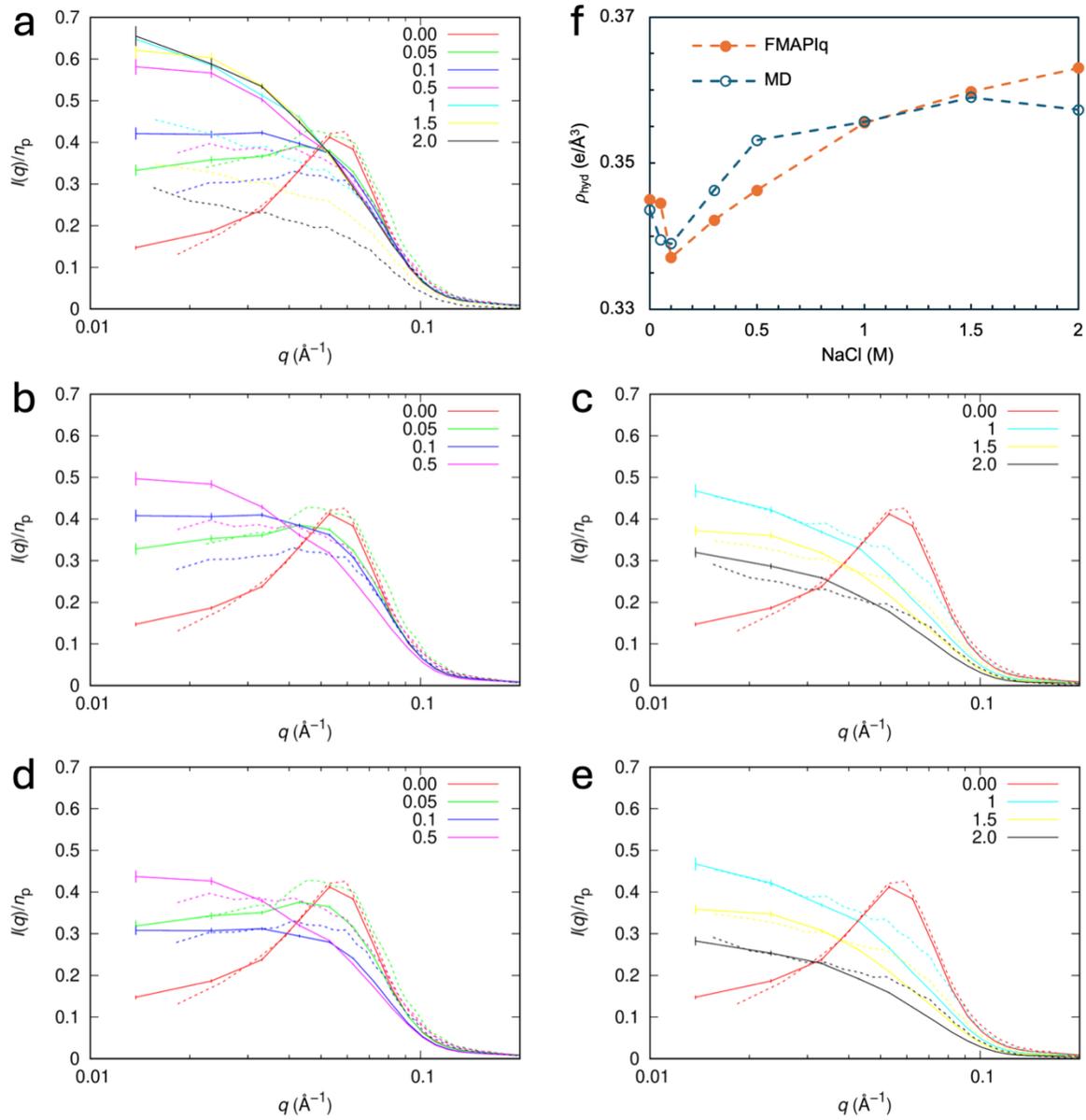

Figure 1

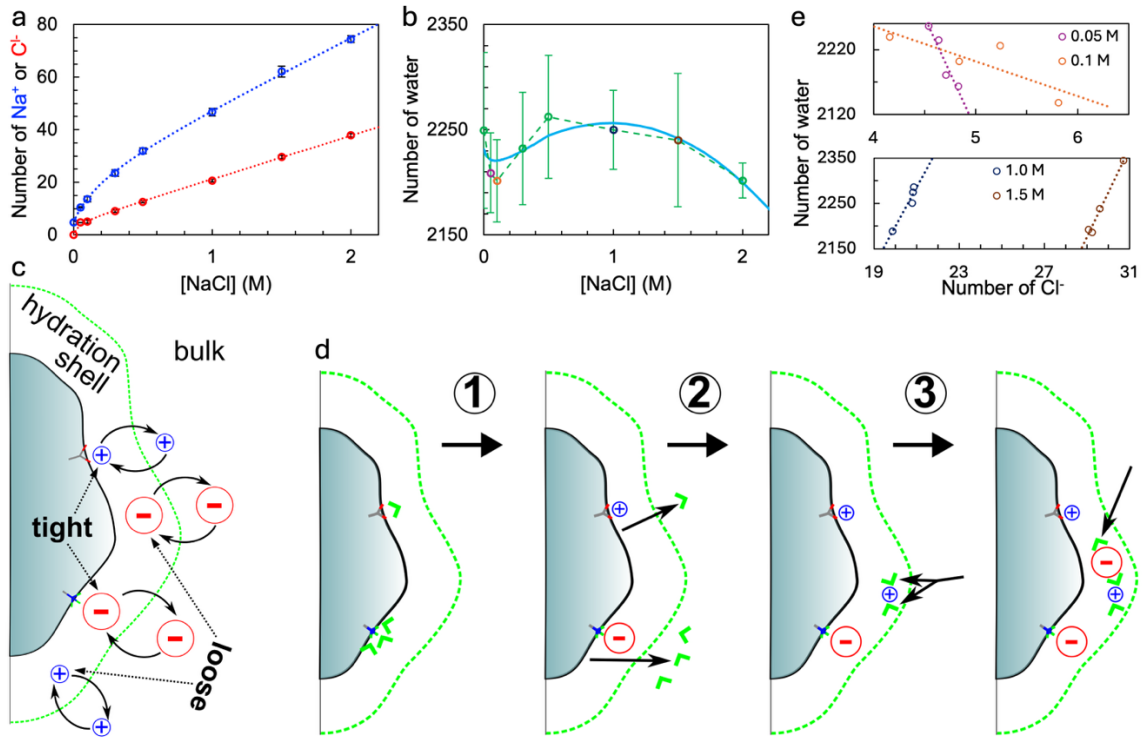

Figure 2

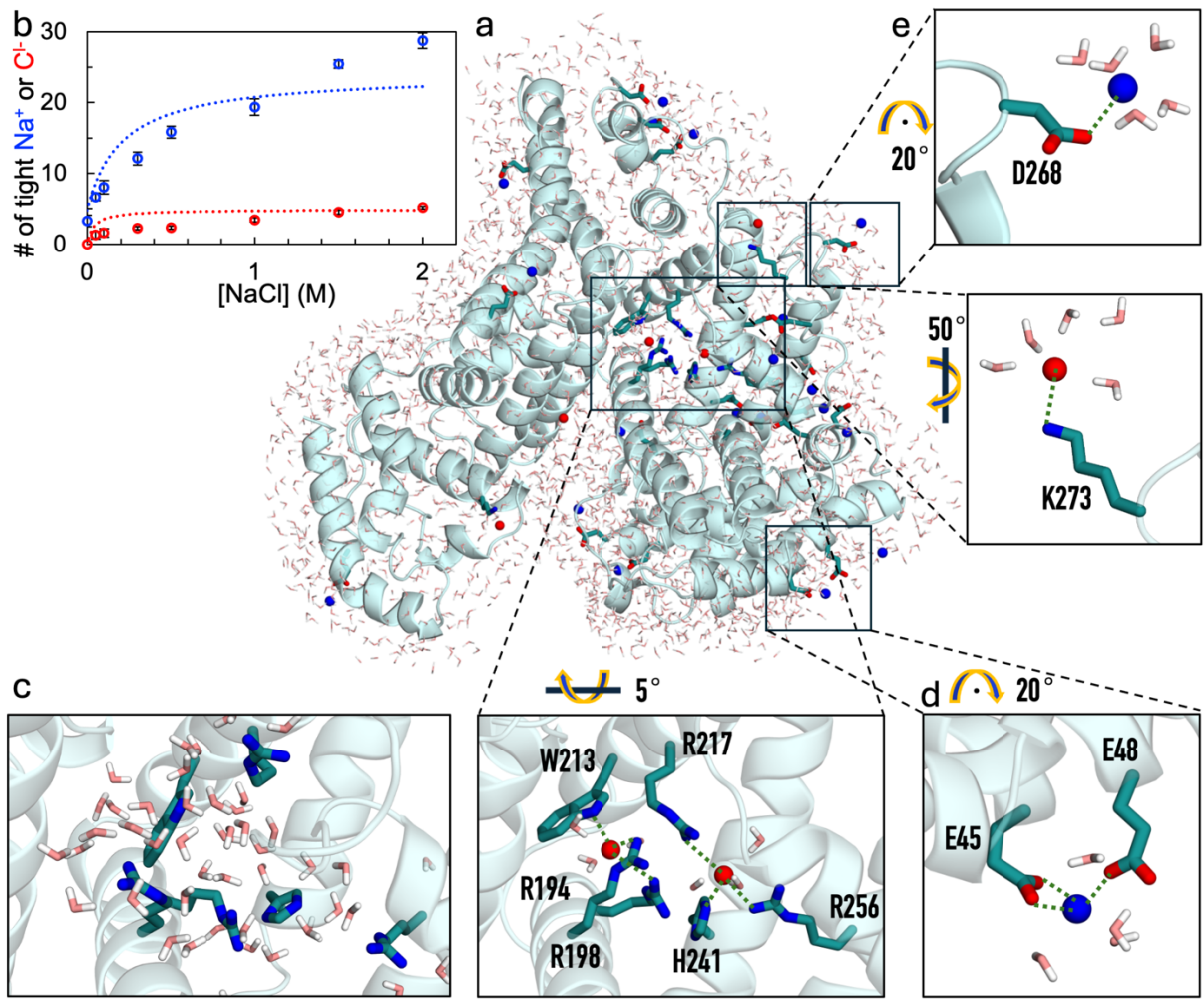

Figure 3

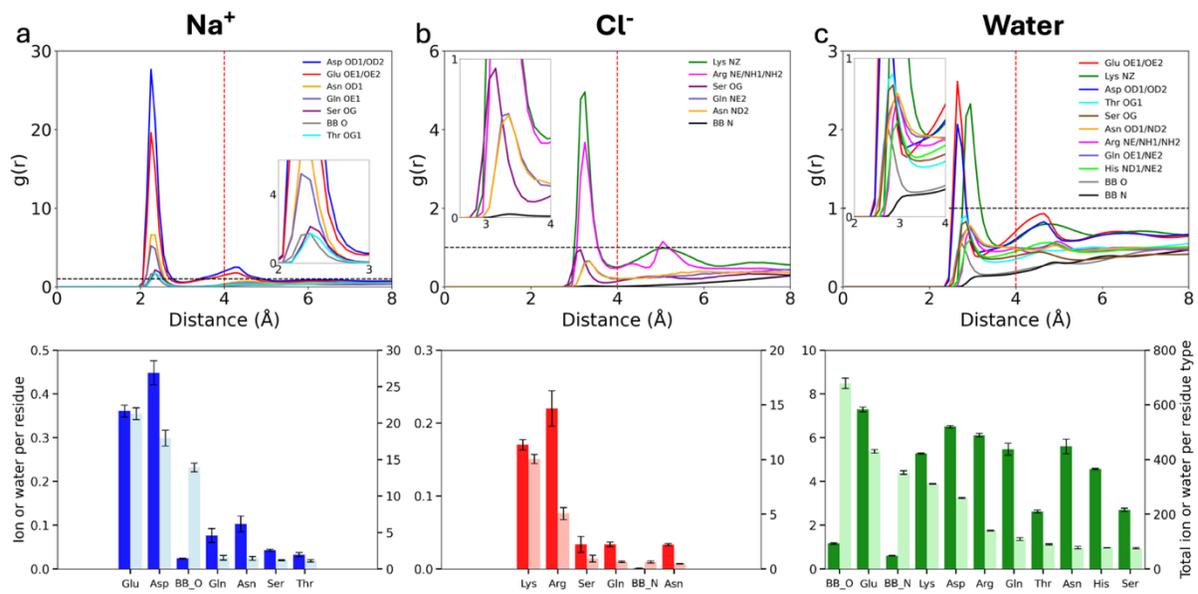

Figure 4